\newcommand{\mybar}[1]{\smash{$\bar{\text{#1}}$}}
\newcommand{\tbar}{\mybar{t}}
\newcommand{\ttbar}{t\tbar\xspace}

\newcommand{\ttW}{t\tbar W\xspace}

\documentclass{SciPost}
\hypersetup{
    colorlinks,
    linkcolor={red!50!black},
    citecolor={blue!50!black},
    urlcolor={blue!80!black}
}

\usepackage[bitstream-charter]{mathdesign}
\usepackage{xspace}
\DeclareSymbolFont{usualmathcal}{OMS}{cmsy}{m}{n}
\DeclareSymbolFontAlphabet{\mathcal}{usualmathcal}

\fancypagestyle{SPstyle}{
\fancyhf{}
\fancyfoot[C]{\textbf{\thepage}}
}

\begin{document}

\pagestyle{SPstyle}

\begin{center}{\Large \textbf{\color{scipostdeepblue}{
Towards a differential \ttW cross section measurement at CMS\\
}}}\end{center}

\begin{center}\textbf{
David M. E. Marckx\textsuperscript{1$\star$}
on behalf of the CMS Collaboration
}\end{center}

\begin{center}
{\bf 1} Ghent University, Ghent, Belgium
\\[\baselineskip]
$\star$ \href{mailto:email1}{\small David.Marcus.E.Marckx@cern.ch}
\end{center}

\definecolor{palegray}{gray}{0.95}
\begin{center}
\colorbox{palegray}{
  \begin{tabular}{rr}
  \begin{minipage}{0.36\textwidth}
    \includegraphics[width=60mm,height=1.5cm]{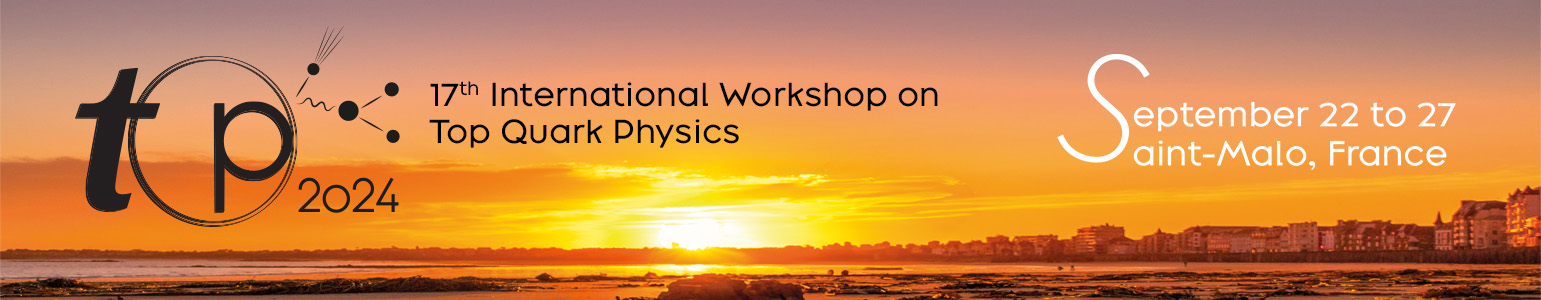}
  \end{minipage}
  &
  \begin{minipage}{0.55\textwidth}
    \begin{center} \hspace{5pt}
    {\it The 17th International Workshop on\\ Top Quark Physics (TOP2024)} \\
    {\it Saint-Malo, France, 22-27 September 2024
    }
    %\doi{10.21468/SciPostPhysProc.?}\\

    \end{center}
  \end{minipage}
\end{tabular}
}
\end{center}

\section*{\color{scipostdeepblue}{Abstract}}
\textbf{\boldmath{%
Top quark pair production in association with a W boson is a rare standard model process that has
proven to be an intriguing puzzle for theorists and experimentalists alike. Recent measurements, performed at $\sqrt{s}$ = 13 TeV, by both the ATLAS and CMS Collaborations at the CERN LHC, find cross section values that are consistently higher than the latest state-of-the-art theory predictions. In this presentation, both experimental and theoretical challenges in the pursuit of a better understanding of this process are discussed.  Furthermore, a framework for a future differential measurement to be performed with the Run 2 CMS data (collected in 2016–2018) is proposed.
}}

\vspace{\baselineskip}

%\noindent\textcolor{white!90!black}{%
%\fbox{\parbox{0.975\linewidth}{%
%\textcolor{white!40!black}{\begin{tabular}{lr}%
%  \begin{minipage}{0.6\textwidth}%
%    {\small Copyright attribution to authors. \newline
%    This work is a submission to SciPost Phys. Proc. \newline
%    License information to appear upon publication. \newline
%    Publication information to appear upon publication.}
%  \end{minipage} & \begin{minipage}{0.4\textwidth}
%    {\small Received Date \newline Accepted Date \newline Published Date}%
%  \end{minipage}
%\end{tabular}}
%}}
%}

%%%%%%%%%% TODO: LINENO
% For convenience during refereeing we turn on line numbers:
%\linenumbers
% You should run LaTeX twice in order for the line numbers to appear.
%%%%%%%%%% END TODO: LINENO

\vspace{10pt}
\noindent\rule{\textwidth}{1pt}
\tableofcontents
\noindent\rule{\textwidth}{1pt}
\vspace{10pt}

\section{Introduction}
\label{sec:intro}

At leading order (LO) in Quantum Chromodynamics (QCD), \ttW production can only proceed via annihilation of an up-type quark with a down-type antiquark or vice versa, where the W boson is radiated from one of the initial state quarks. Only at Next-to-LO (NLO), additional production channels are available for \ttW production, initiated by a quark-gluon interaction and containing an additional jet in the final state. This seemingly simple process has however proven to be extremely challenging to model, and recent measurements performed by both the ATLAS and CMS Collaborations at the CERN LHC show a small, but consistent, tension with the latest theory predictions~\cite{atlasttw,CMS:2022tkv}.

\vspace{3mm}

\noindent In this contribution, we discuss the many challenges we face in performing a \ttW measurement. We then report the inclusive cross-section measurement from Ref.~\cite{CMS:2022tkv}, which is based on 138 fb$^{-1}$ of data recorded with the CMS experiment~\cite{TheCMSCollaboration_2008,CMS:Detector-2024} in 2016--2018. Lastly, we explore the possibility and methodology of performing a differential cross-section measurement.

\section{Challenges in \ttW}
\subsection{Challenges for theoretical calculations}
\label{sec:theory}
\begin{figure}[h!]
	\centering
	\includegraphics[height=0.18\linewidth]{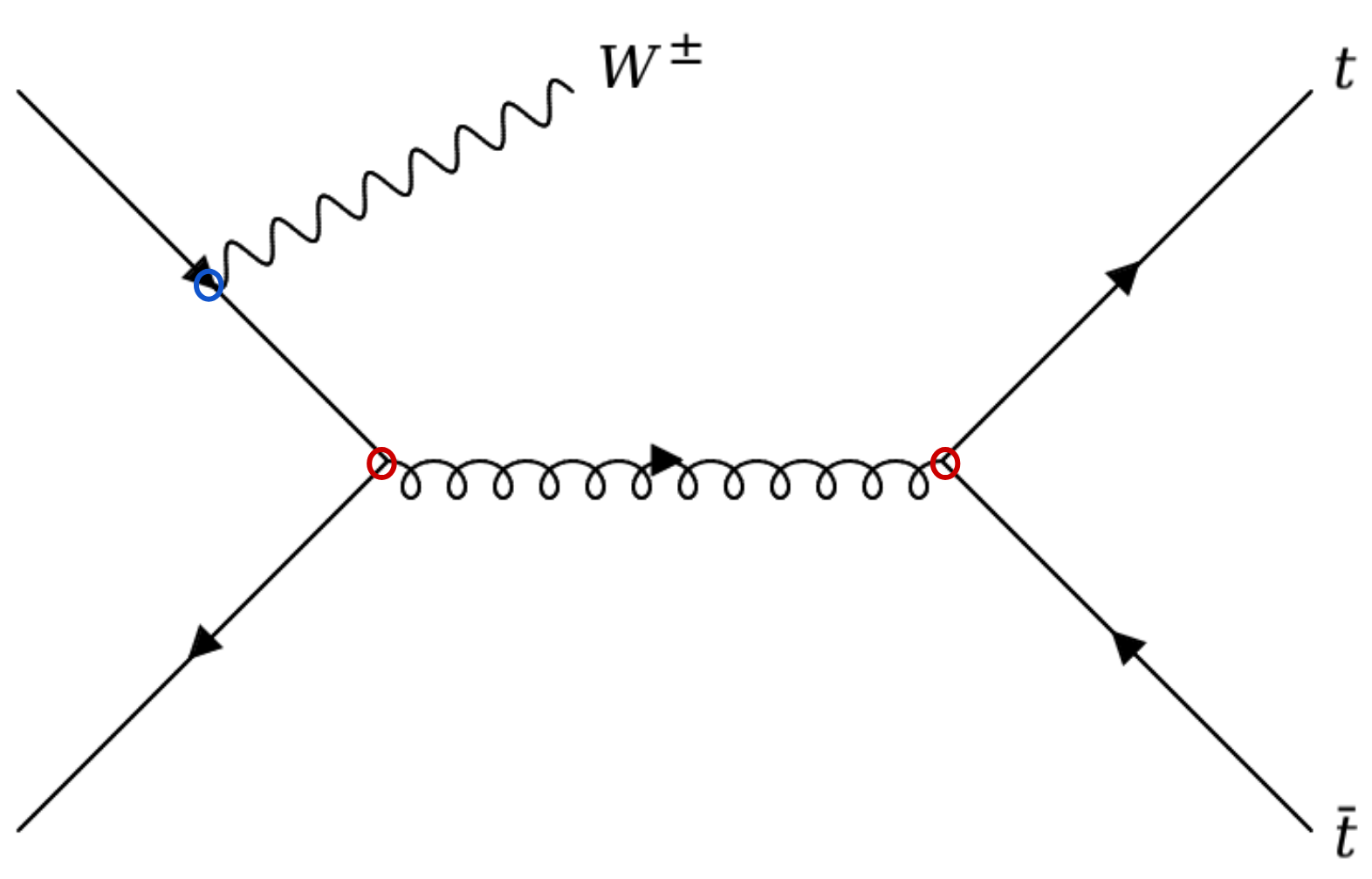}
	\includegraphics[height=0.18\linewidth]{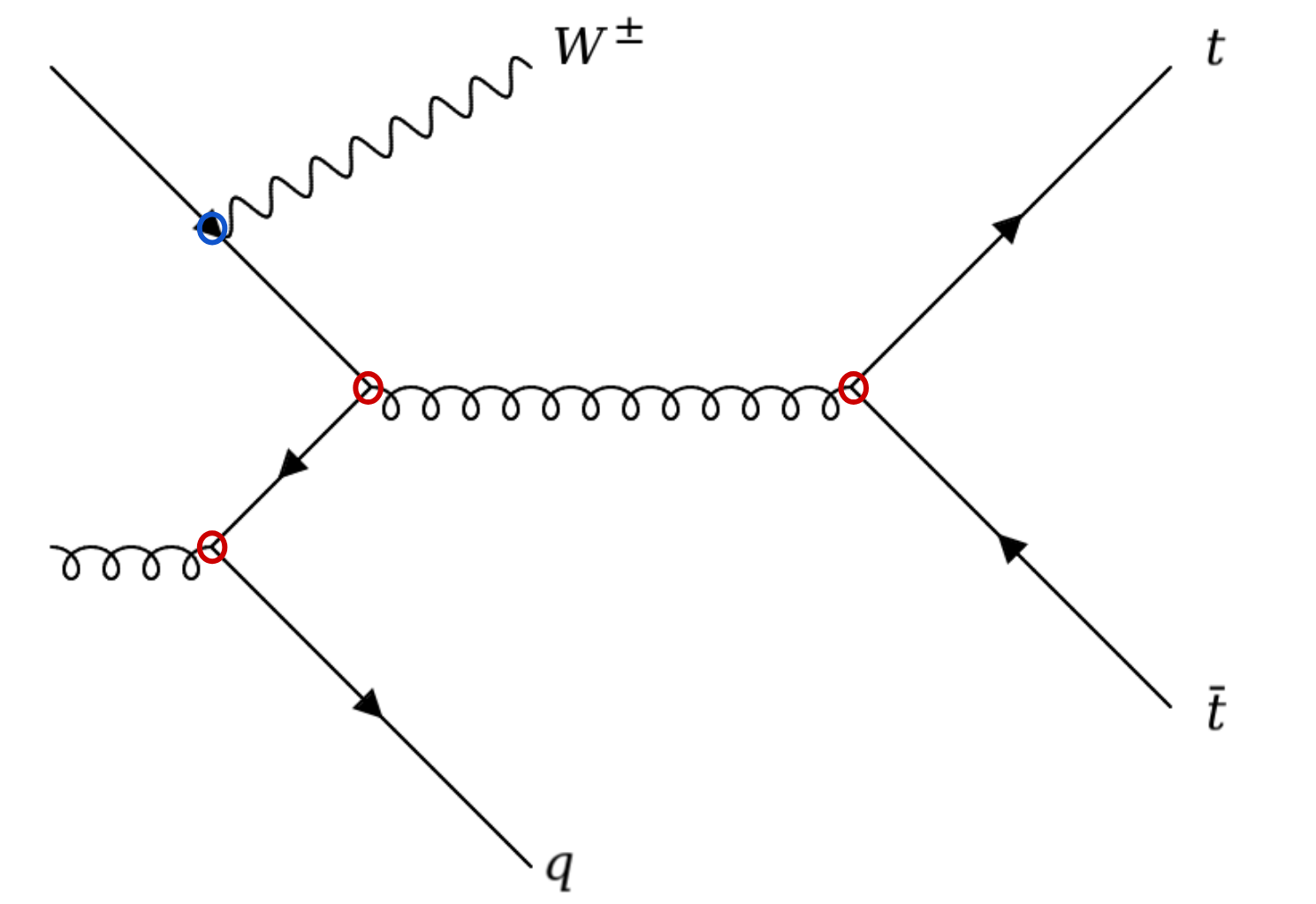}
    \includegraphics[height=0.21\linewidth]{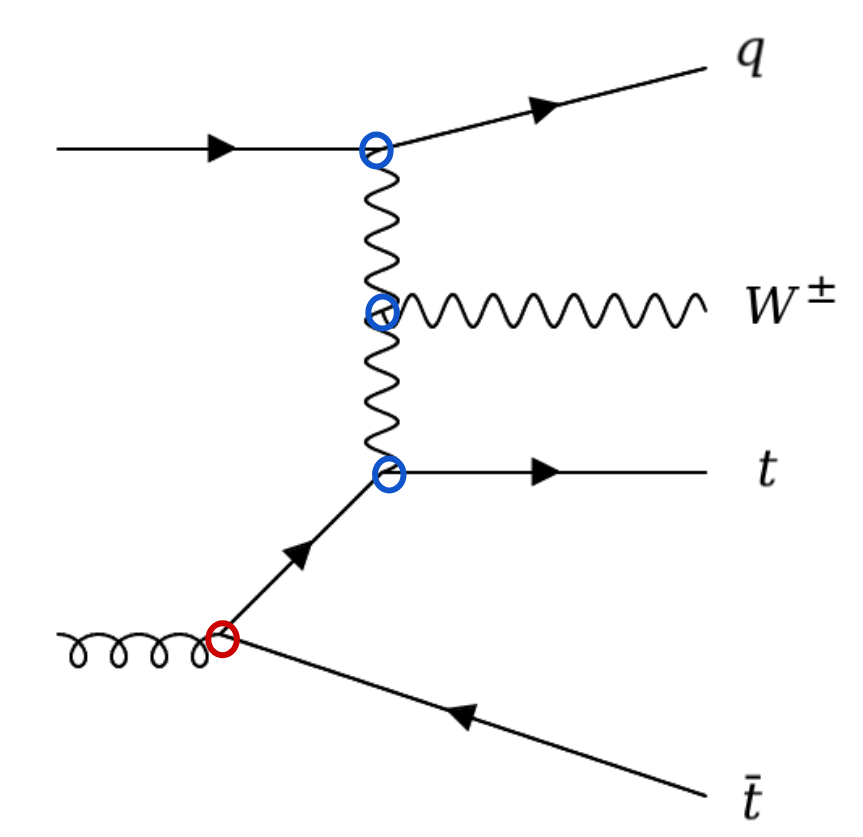}
	\caption{Feynman diagrams contributing to \ttW production. (left) The LO diagram contributing to \ttW production. (middle) A NLO QCD diagram that has a gluon in the initial state. (right) A t-W scattering diagram that contributes to the subleading EWK NLO term. The red (blue) circles show the QCD (EWK) couplings.}
	\label{diagrams}
\end{figure}

As there are only two diagrams contributing to \ttW production at LO in QCD, going to higher order calculations and calculations with more electroweak (EWK) contributions allows for many additional diagrams to contribute to the \ttW cross section. These extra diagrams, shown in Fig.~\ref{diagrams}, give significant corrections to the overall cross section and modify significantly the predicted event kinematics. For instance, the NLO QCD term opens up the quark-gluon induced channel. With a much larger gluon proton parton distribution function (pdf) contribution, this term modifies the overall cross section with a k-factor of roughly 1.40, while one would expect an effect of around 10\% from naive power counting~\cite{Frederix:2017wme}.

\noindent As an even more extreme example, one can consider the subleading EWK NLO contributions. From naive power counting assumptions, one can expect modifications of the total cross section to be of the order 0.1\%. But the t-W scattering diagrams, highlighted in Fig.~\ref{diagrams}, actually result in a k-factor of 1.12~\cite{Frederix:2017wme}. These larger than expected contributions show the importance of including higher-order contributions in our theoretical calculations. Many of these higher-order contributions are however extremely difficult to compute. in order to compute the full NNLO QCD contributions to \ttW production, the tree-level amplitudes and one-loop scattering amplitudes are already available but the two-loop virtual corrections are not yet known. 

\subsubsection{State-of-the-art}
\begin{figure}[h!]
	\centering
    \raisebox{-0.5\height}{\includegraphics[height=0.2\linewidth]{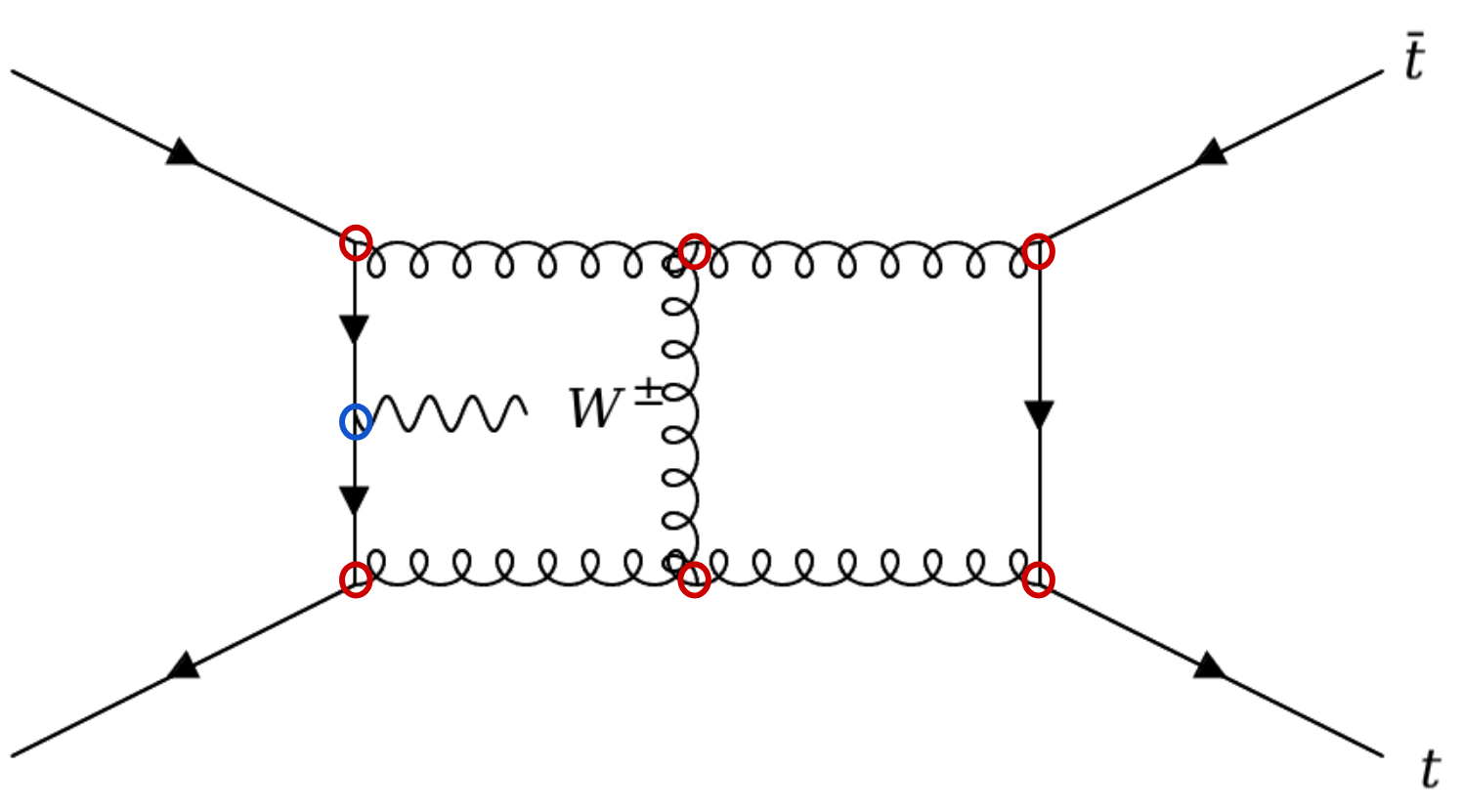}}
	\raisebox{-0.5\height}{\includegraphics[height=0.45\linewidth]{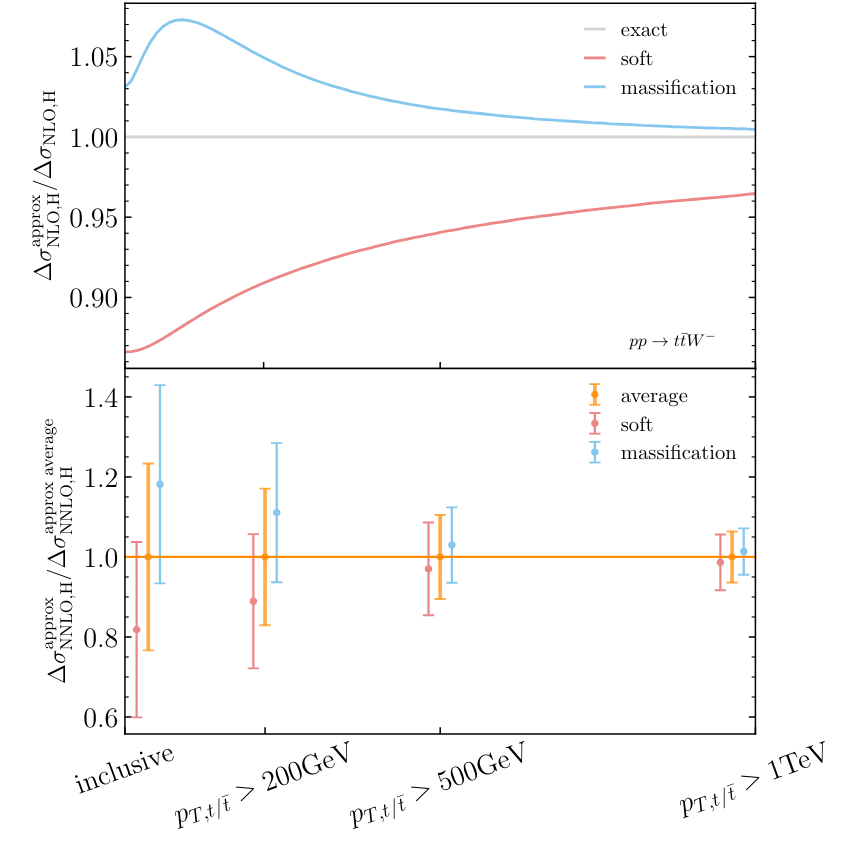}}
	\caption{(left) Double-loop diagram, for which the amplitude is yet unknown. (right) The convergence of both approximations used to estimate the double-loop contribution. The top pannel shows the convergence to the exact calculation at NLO QCD. The bottem pannel shows the convergence at NNLO~\cite{Buonocore_2023}.}%, where the exact calculation is of course not available.}
	\label{approx}
\end{figure}

Currently the state-of-the-art calculations can only approximatively calculate two-loop virtual contributions to the full NNLO QCD cross section of \ttW. The first approximation calculates the amplitude in the ultra-relativistic limit with a massification method. On a simplistic level, with $Q>>m_t$, we can get an amplitude with "massless" top quarks, which is known from NNLO W+light jets calculations. The other approximation assumes the radiation of a soft W boson. In this limit the amplitude can be factorised into the virtual amplitude of \ttbar production and a soft W emission. As we do know that these corrections are significant, we can not yet trust these approximations for kinematic distributions at NNLO. Therefore, only the inclusive cross-section of 745.3 fb $\pm 6.7\%$ (scale) $\pm1.8\%$ (approx.) is provided by the authors~\cite{Buonocore_2023}. The two approximations at NNLO and their convergence to the true value for NLO calculations are shown in figure~\ref{approx}.

\begin{figure}[b!]
	\centering
    \includegraphics[height=0.4\linewidth]{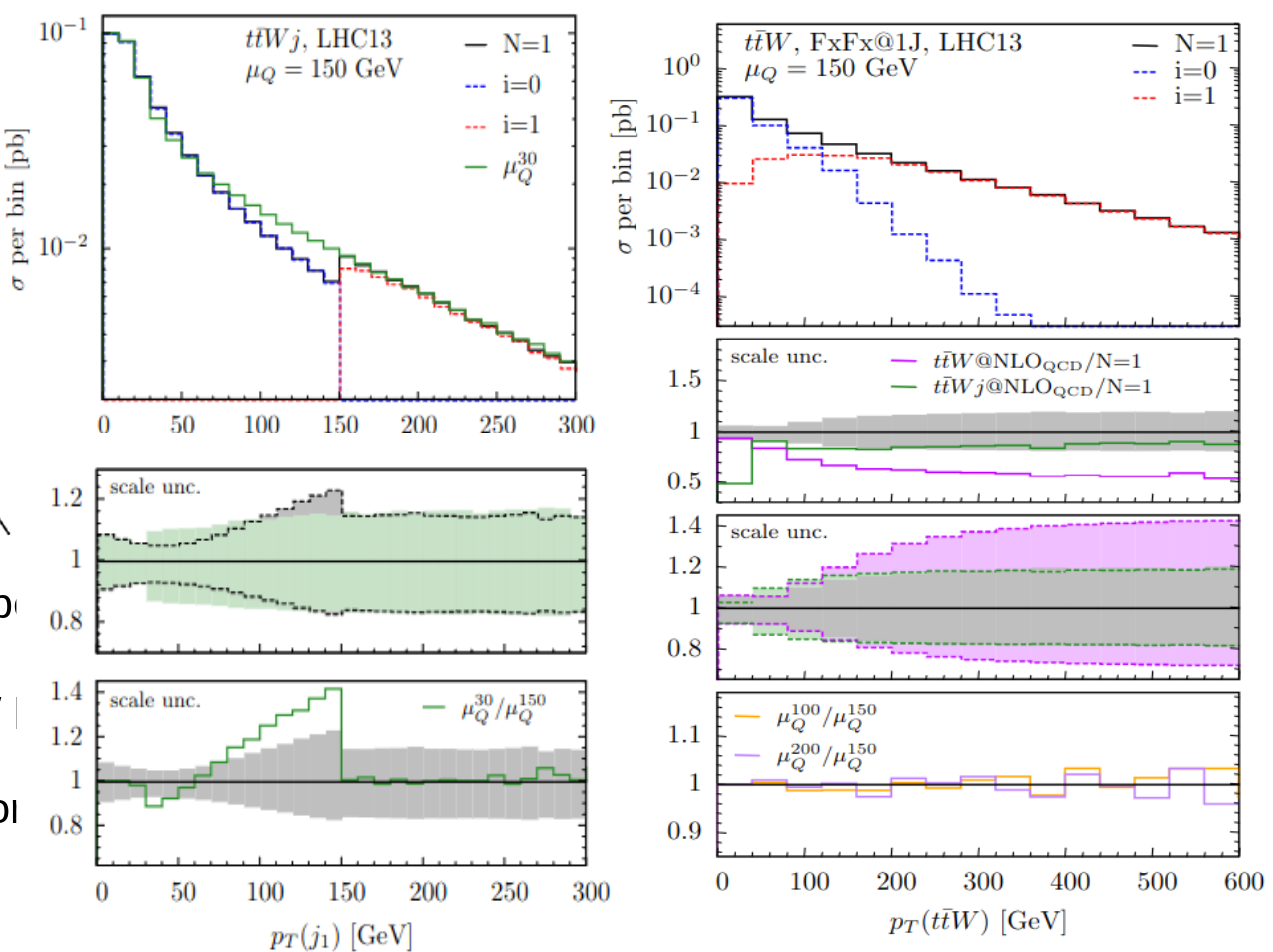}
	\caption{(left) Merging scale dependence for \ttW production with the old FxFx@1j merging scheme~\cite{Frederix_2012}. (right) Merging scale dependence for the new FxFx scheme~\cite{Frederix_2021}.}
	\label{merging}
\end{figure}

\noindent Apart from higher-order calculations, there have been numerous other improvements in the modelling of the \ttW process. In this presentation we also highlighted an improved parton-jet merging scheme developed by the original FxFx authors~\cite{Frederix_2012}. For this improved scheme, it was realised that the emission of a W from final-state extra partons protects these partons from infrared (IR) divergences. Kinematically identifying these "electroweak" jets allows one to exclude them from the merging procedure and hence keep these jets from the matrix element even below the merging scale $Q_{\text{cut}}$. Having jets described by the matrix element below the merging scale could improve the modelling of these extra emissions~\cite{Frederix_2021} and lowers the merging scale dependence, as shown in Fig.~\ref{merging}.

\subsection{Challenges for experimental measurements}
\label{sec:experiment}
\begin{figure}[h!]
	\centering
    \includegraphics[width=0.4\linewidth]{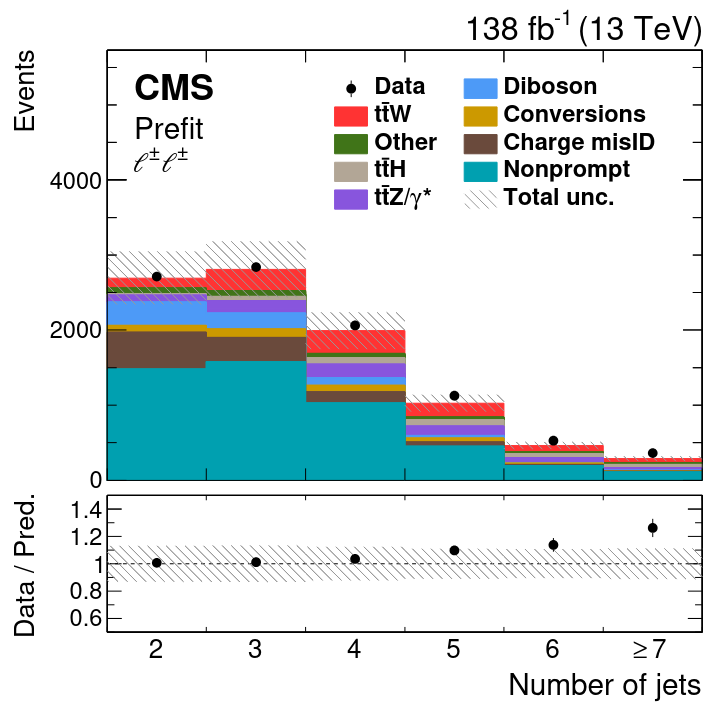}
    \includegraphics[width=0.4\linewidth]{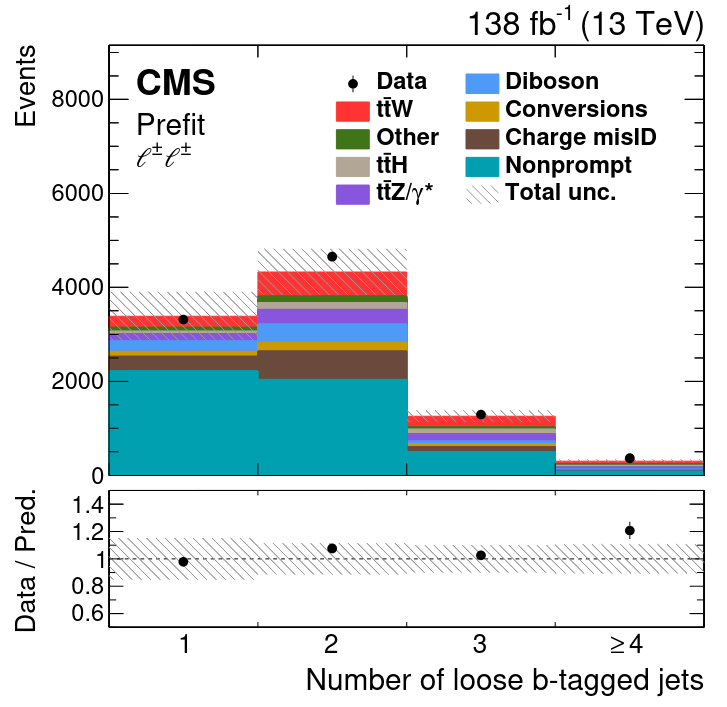}
	\caption{Number of jets and number of b-tagged jets in the 2l SS signal region of the CMS inclusive \ttW analysis~\cite{CMS:2022tkv}.}
	\label{jets}
\end{figure}

While \ttW has proven to be a big challenge for the modelling community, it also poses considerable issues for experimental analyses. In experimental measurements, most sensitivity can be reached in a 2 lepton (2l) same-sign (SS) signal region. A high number of (b-)jets allows us to exclude many background processes that are not top-related. Furthermore requiring two SS leptons excludes a large amount of other top-related processes such as top pair production and top pair production in association with a Higgs, Z boson or $\gamma$. As can be seen in Figure~\ref{jets}, however, one still does not reach a significant purity. Next to remnants of the processes mentioned above, significant background contributions come from reducible backgrounds: non-prompt leptons, charge-misidentified electrons and photon conversions in the detector material. 

\noindent To control all the reducible backgrounds, one needs to rely on a good lepton identification (ID) algorithm, which is able to identify prompt leptons with a high probability. Good b-tagging performance is furthermore important. As the reducible backgrounds are known to be ill-described by simulation and many backgrounds are ill-described in this specific phase space region, one needs to rely on many data-driven background estimation methods and control regions to better understand these backgrounds.

\section{Cross section measurement}
\label{sec:inclusive}
\begin{figure}[h!]
	\centering
    \includegraphics[width=0.8\linewidth]{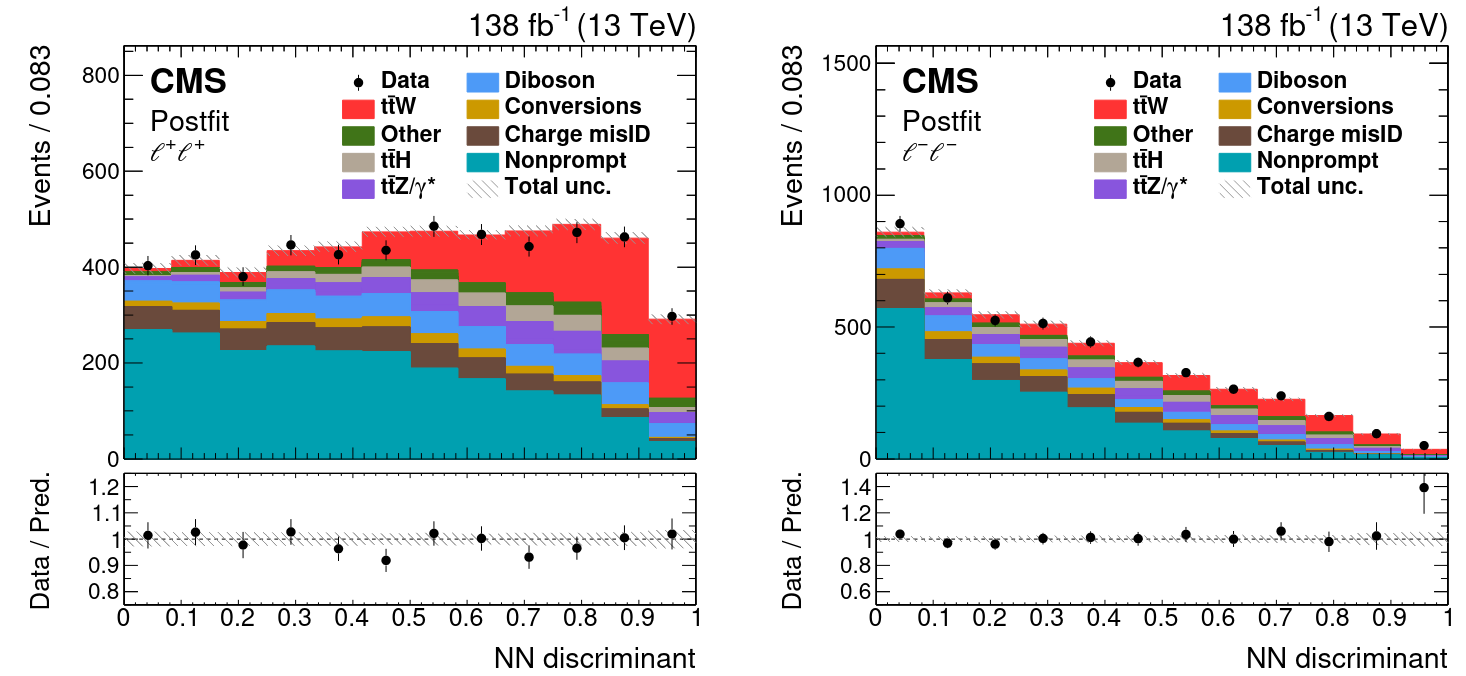}
	\caption{The neural network distribution in the dilepton signal region for leptons with positive (left) or negative (right) charge~\cite{CMS:2022tkv}.}
	\label{bdt}
\end{figure}

The inclusive cross-section measurement by the CMS collaboration~\cite{CMS:2022tkv} is designed with an emphasis on maximizing the signal acceptance. As can be seen in Figure~\ref{jets}, the signal purity could for instance be increased by requiring more (b-)jets, having a tighter lepton ID or a tighter b-tagging requirement. Signal purity in the measurement is achieved by employing a neural network that is trained to separate \ttW events from other background processes. The signal region can furthermore be split based on the charge of the final state leptons, as seen in Fig.~\ref{bdt}. This way the charge asymmetry that is present in \ttW production due to the asymmetry in the proton pdf, which is much less visible in processes that are predominantly gluon induced at leading order, is fully used in constraining the background.

\subsection{Towards a differential measurement}
Expanding the framework used in the inclusive analysis towards a differential measurement, many of the same ingredients can be used. As many uncertainties and backgrounds are determined in situ, the statistical model used in the analysis is of extreme importance. It is therefore beneficial to encode the unfolding procedure of the differential measurement into the statistical model. For this one can use a Maximum-Likelihood based unfolding strategy. The signal process is split based on MC truth level information, this way the response matrix migrations are encoded into the Maximum likelihood fit and migrations are controlled by the uncertainty profiling procedure. To diagonalise the unfolding as much as possible, the NN distribution is split at the reconstructed level based on the variable of interest.

\section{Conclusion}
Challenges to improving our understanding of the \ttW process have been presented in this talk. Some current state-of-the-art theoretical results have been highlighted and the measurement of the inclusive \ttW cross-section by the CMS collaboration has been discussed. An outline of the strategy for an ongoing differential cross-section measurement analysis has also been introduced, which will help theorists provide more feedback for their calculations of the \ttW process.

\section*{Acknowledgements}
The author of these proceedings would like to express his sincere gratitude to J. Vandenbroeck and N. Van Den Bossche for their help in the writing of these proceedings. Their expertise and support were instrumental in shaping this work.

\iffalse
\begin{appendix}
\numberwithin{equation}{section}

\section{First appendix}
Add material which is better left outside the main text in a series of Appendices labeled by capital letters.

\section{About references}
Your references should start with the comma-separated author list (initials + last name), the publication title in italics, the journal reference with volume in bold, start page number, publication year in parenthesis, completed by the DOI link (linking must be implemented before publication). If using BiBTeX, please use the style files provided  on \url{https://scipost.org/submissions/author_guidelines}. If you are using our LaTeX template, simply add
\begin{verbatim}
\bibliography{your_bibtex_file}
\end{verbatim}
at the end of your document. If you are not using our LaTeX template, please still use our bibstyle as
\begin{verbatim}
\bibliographystyle{SciPost_bibstyle}
\end{verbatim}
in order to simplify the production of your paper.
\end{appendix}
\fi

% Use your bibtex library, formatted by the SciPost style file.
\bibliography{SciPost_Example_BiBTeX_File.bib}

\end{document}